\documentclass[reprint,pra, longbibliography,superscriptaddress,amsmath,amssymb,aps]{revtex4-2}

\usepackage{graphicx}% Include figure files
\usepackage{dcolumn}% Align table columns on decimal point
\usepackage{bm}% bold math
\usepackage[T1]{fontenc}
\usepackage{siunitx}
\usepackage{chemformula}
\usepackage{CJKutf8}
\usepackage{xcolor}

\let\vaccent=\v % rename builtin command \v{} to \vaccent{}
\renewcommand{\v}[1]{\ensuremath{\mathbf{#1}}} % for vectors
 
\newcommand{\abs}[1]{\left| #1 \right|} % for absolute value
\newboolean{ShowComments}
\setboolean{ShowComments}{true}  % change to true to show remaining colorful author comments. 
\ifthenelse{\boolean{ShowComments}}%
	{
		\newcommand{\ColorComment}[3]{%
				{\colorbox{#1}{\color{white}   \textsf{\textbf{#2}}} \textcolor{#1}{#3}}}%  Colorful box, initials, phrase 
		\newcommand{\nyacite}[1]{[#1]}% not yet a cite
	}%
	{
		\newcommand{\ColorComment}[3]{}%  Do nothing at all
		\newcommand{\nyacite}[1]{}% not yet a cite -- do nothing
	}%
\definecolor{newcolor}{rgb}{0.5,0,0}

\usepackage{cancel}

\begin{document}

\title{
Valley Splitting Correlations Across a Silicon Quantum Well Containing Germanium}

\author{Jonathan C. Marcks}
\altaffiliation{Corresponding author: Jonathan C. Marcks, jmarcks@anl.gov}
\affiliation{Q-NEXT, Argonne National Laboratory, Lemont, Illinois, 60439, United States}
\affiliation{Materials Science Division, Argonne National Laboratory, Lemont, Illinois, 60439, United States}
\affiliation{Pritzker School of Molecular Engineering, University of Chicago, Chicago, Illinois, 60637, United States}

\author{Emily Eagen}
\affiliation{Department of Physics, University of Wisconsin-Madison, Madison, Wisconsin, 53706, United States}

\author{Emma C. Brann}
\affiliation{Department of Physics, University of Wisconsin-Madison, Madison, Wisconsin, 53706, United States}

\author{Merritt P. Losert}
\affiliation{Department of Physics, University of Wisconsin-Madison, Madison, Wisconsin, 53706, United States}

\author{Talise Oh}
\affiliation{Department of Physics, University of Wisconsin-Madison, Madison, Wisconsin, 53706, United States}

\author{J. Reily}
\affiliation{Department of Physics, University of Wisconsin-Madison, Madison, Wisconsin, 53706, United States}
\affiliation{Materials Science Division, Argonne National Laboratory, Lemont, Illinois, 60439, United States}

\author{Christopher S. Wang}
\affiliation{James Franck Institute and Department of Physics, University of Chicago, Chicago, Illinois, 60637, United States}
\affiliation{Q-NEXT, Argonne National Laboratory, Lemont, Illinois, 60439, United States}

\author{Daniel Keith}
\affiliation{Intel Technology Research, Intel Corporation, Hillsboro, Oregon, 97124, United States}

\author{Fahd A. Mohiyaddin}
\affiliation{Intel Technology Research, Intel Corporation, Hillsboro, Oregon, 97124, United States}

\author{Florian Luthi}
\affiliation{Intel Technology Research, Intel Corporation, Hillsboro, Oregon, 97124, United States}

\author{Matthew J. Curry}
\affiliation{Intel Technology Research, Intel Corporation, Hillsboro, Oregon, 97124, United States}

\author{Jiefei Zhang}
\affiliation{Q-NEXT, Argonne National Laboratory, Lemont, Illinois, 60439, United States}
\affiliation{Materials Science Division, Argonne National Laboratory, Lemont, Illinois, 60439, United States}

\author{F. Joseph Heremans}
\affiliation{Q-NEXT, Argonne National Laboratory, Lemont, Illinois, 60439, United States}
\affiliation{Pritzker School of Molecular Engineering, University of Chicago, Chicago, Illinois, 60637, United States}
\affiliation{Materials Science Division, Argonne National Laboratory, Lemont, Illinois, 60439, United States}

\author{Mark Friesen}
\affiliation{Department of Physics, University of Wisconsin-Madison, Madison, Wisconsin, 53706, United States}

\author{M. A. Eriksson}
\affiliation{Department of Physics, University of Wisconsin-Madison, Madison, Wisconsin, 53706, United States}

\begin{abstract}
Quantum dots in SiGe/Si/SiGe heterostructures host coherent electron spin qubits, which are promising for future quantum computers. The silicon quantum well hosts near-degenerate electron valley states, creating a low-lying excited state that is known to reduce spin qubit readout and control fidelity. The valley energy splitting is dominated by the microscopic disorder in the SiGe alloy and at the Si/SiGe interfaces, and while Si devices are compatible with large-scale semiconductor manufacturing, achieving a uniformly large valley splitting energy across a many-qubit device spanning mesoscopic distances is an outstanding challenge. In this work we study valley splitting variations in a 1D quantum dot array, formed in a \ch{Si_{0.972}Ge_{0.028}} quantum well, manufactured by Intel. We observe correlations in valley splitting, at both sub-\SI{100}{\nano\meter} (single gate) and $>\SI{1}{\micro\meter}$ (device) lengthscales, that are consistent with alloy disorder-dominated theory and simulation. Our results develop the mesoscopic understanding of Si/SiGe heterostructures necessary for scalable device design.
\end{abstract}

\maketitle

\section{Introduction}
The spin of an electron is a natural physical system for encoding quantum information~\cite{loss_quantum_1998,awschalom_quantum_2013}. Gate-defined semiconductor quantum dots, where single or few electrons are trapped in materials like silicon~\cite{eriksson_spin-based_2004,simmons_tunable_2011,zwanenburg_silicon_2013,burkard_semiconductor_2023}, are likewise a promising platform for building quantum computers out of electron spin qubits~\cite{morton_embracing_2011,vandersypen_interfacing_2017,philips_universal_2022,weinstein_universal_2023}. Si, owing to its nuclear spin-free isotope and low spin-orbit coupling~\cite{zwanenburg_silicon_2013}, can host coherent spin qubits with gate fidelities in the regime needed for quantum error correction~\cite{fowler_high-threshold_2009,veldhorst_addressable_2014,yoneda_quantum-dot_2018,noiri_fast_2022}, likely necessary for realizing useful quantum computation. Recent demonstrations of industrial manufacturing of Si quantum dot devices~\cite{zwerver_qubits_2022,neyens_probing_2024,elsayed_low_2024,steinacker_industry-compatible_2025,george_12-spin-qubit_2025,huckemann_industrially_2024} bring the expertise of the semiconductor industry to the challenge of producing uniform, low-noise, scalable quantum dot processors~\cite{stuyck_CMOS_2024}. Ideally, the spin qubit energy levels in such processors are the lowest-energy excitations in the system and are well-isolated from any excited states~\cite{divincenzo_physical_2000}. However, the indirect bandgap and six-fold crystal symmetry of silicon give rise to six emergent electronic valley states at the degenerate conduction band minima~\cite{phillips_band_1962,schaffler_high-mobility_1997}. In strained Si quantum wells embedded in SiGe heterostructures---a common quantum dot platform---the in-plane strain breaks this degeneracy and lifts the four in-plane valleys by hundreds of meV~\cite{schaffler_high-mobility_1997}, but the out-of-plane confinement of the electron wavefunction typically splits the two out-of-plane valleys by only \SIrange{10}{200}{\micro\eV}~\cite{mi_high-resolution_2017,hollmann_large_2020,chen_detuning_2021,dodson_how_2022,paquelet_wuetz_atomic_2022,volmer_mapping_2024}. The low-lying excited state degrades qubit control and state preparation and measurement (SPAM) fidelities and is a major roadblock to building quantum dot processors~\cite{culcer_realizing_2009,friesen_theory_2010,gamble_disorder-induced_2013,burkard_semiconductor_2023}.

Recent theoretical work has demonstrated that alloy disorder---the random distribution of Ge atoms in the SiGe alloy that modulates the quantum well potential and breaks the degeneracy between the out-of-plane valley states---is the dominant factor limiting valley splitting in the majority of experiments to $<\SI{100}{\micro\eV}$~\cite{paquelet_wuetz_atomic_2022,losert_practical_2023}. This has led to proposals to uniformly raise the distribution of valley splitting in a given heterostructure by engineering the quantum well potential~\cite{degli_esposti_low_2024}, notably through the incorporation of Ge in the well in addition to the surrounding alloy~\cite{feng_enhanced_2022,mcjunkin_sige_2022,losert_practical_2023,losert_strategies_2024}. Recent investigations have demonstrated a trade-off between Ge-induced valley splitting enhancements and the commensurate increase in disorder, as measured by mobility and percolation density~\cite{stehouwer_engineering_2025}. At few-percent concentrations---or through more sophisticated incorporation profiles~\cite{feng_enhanced_2022,mcjunkin_sige_2022}---the well still confines the electron~\cite{schaffler_high-mobility_1997} while increasing the coupling potential between the valleys on average~\cite{losert_practical_2023}. However, these strategies still leave regions of material where the electron will experience low valley splitting~\cite{losert_practical_2023,volmer_mapping_2024,losert_strategies_2024}, which would reduce the yield of manufactured quantum dots and impact electron shuttling schemes~\cite{kunne_spinbus_2024}. For the application of these strategies to future heterostructure design, it is necessary to investigate the inherent variations in the valley splitting of state-of-the-art engineered quantum dot devices.

In this work, we observe spatial correlations in valley splitting energy, using the electron wavefunction as a probe of underlying material disorder along the \SI{1.3}{\micro\meter}-long Ge-containing (\ch{Si_{0.972}Ge_{0.028}}) channel of an industrially fabricated Intel Tunnel Falls device~\cite{george_12-spin-qubit_2025}. We reveal fast changes with nm-scale resolution underneath individual quantum dot gates arising from local alloy disorder, and then we identify correlations in valley splitting along the quantum channel that arise from material fluctuations. After describing the electrostatic and valley device modeling, we explain the detuning axis pulsed spectroscopy (DAPS)~\cite{chen_detuning_2021} technique for measuring valley splitting in a double quantum dot configuration with a continuously variable position. We present gate- and device-scale valley splitting measurements, finding correlations consistent with alloy disorder-dominated (ADD) valley splitting~\cite{losert_practical_2023}. This study reveals the emergence of non-trivial variations in single electron level structure arising from the inherent inhomogeneity of semiconductor heterostructures, and will enable both improved design of Si/SiGe materials and improved modeling of valley-induced qubit infidelity.

\section{Results}
\subsection{Device modeling}
\begin{figure}
    \centering
    \includegraphics{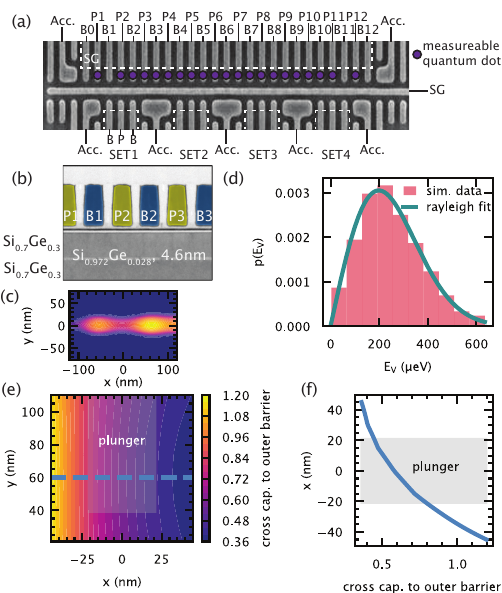}
    \caption{\bf Ge-containing quantum well devices and modeling. \rm (a) Top-down SEM of quantum dot and single electron transistor (SET), plunger (P), barrier (B), accumulation (Acc.), and screening (SG) gates in an Intel Tunnel Falls device nominally identical to the device studied in this paper. Screening gates buried underneath the qubit and SET finger gates are outlined in white. The symmetry between plunger and barrier gates allows formation of 21 separate quantum dots (purple circles) in the top channel. (b) Side-on view of a SiGe heterostructure with a \SI{4.6}{\nano\meter}-thick Si quantum well and finger gates. The well in this work is \ch{Si_{0.972}Ge_{0.028}} (the image is of a pure Si well). (c) Electron density in a tuned up double dot demonstrating dot elongation along the 1D channel ($x$). (d) Distribution of simulated valley splitting energies $E_V$ for the device heterostructure according to the alloy disorder-dominated theory in Ref.~\cite{losert_practical_2023}. Overlaid fit is to a Rayleigh distribution. (e) Simulated cross capacitance of an electron under a plunger gate to the outer barrier gate. (f) Electron center position $x$ versus simulated cross capacitance along the linecut in (e), used to calibrated dot position $x_{dot}$ in Fig.~\ref{fig:fig3}.}
    \label{fig:fig1}
\end{figure}

We study a device (Intel Tunnel Falls) nominally identical to that shown in Fig.~\ref{fig:fig1}(a,b), designed to host up to 12 quantum dots. The symmetry between the barrier and plunger gates in Tunnel Falls devices allows formation of quantum dots under either plunger or barrier gates, effectively swapping the gate roles. 21 different dots are possible, shown as purple circles in Fig.~\ref{fig:fig1}(a), under gates P1, P2-P11 (including B gates), and P12. Dots cannot be formed under B0 or B12 because they are next to the accumulation gates, and while double dot pairs B1-B2 and B10-B11 are possible, the symmetry with the SET1 and SET4 sensors precludes their measurement. With a gate pitch of \SI{60}{\nano\meter}, the active area of the device spans \SI{1.3}{\micro\meter} of material.

Fig.~\ref{fig:fig1}(b) is a cross-sectional TEM of a heterostructure with the same well and gate geometry as the device studied in this paper, with a \SI{4.6}{\nano\meter}-thick, \SI{2.8}{\%}-Ge Si quantum well (QW) in bulk \ch{Si_{0.7}Ge_{0.3}} grown on a Si substrate (the figure image is of a pure Si well). Details of the industrial fabrication process can be found in Ref.~\cite{george_12-spin-qubit_2025}. The QW parameters are chosen to optimize for large average valley splitting according to recent theory proposals and experimental validation~\cite{chen_detuning_2021,paquelet_wuetz_atomic_2022,losert_practical_2023,watson_2025}.

Simulation of the electron density under normal operating voltages in a double quantum dot, performed in MaSQE~\cite{anderson_high-precision_2022}, is shown in Fig.~\ref{fig:fig1}(c), demonstrating the dot is elliptical, with a long axis along the dot channel ($x$ direction). This justifies the assignment of the first excited orbital state in proceeding measurements as an $x$-orbital. A description of the MaSQE approach and orbital state calculations is in SI Sec.~S1. Fig.~\ref{fig:fig1}(d) calculates 1000 valley splitting values assuming the heterostructure in Fig.~\ref{fig:fig1}(b) and an ADD theory~\cite{losert_practical_2023}. The simulated data shows an average valley splitting of \SI{200}{\micro\eV} and fits to the expected Rayleigh distribution, providing a theoretical reference for valley splitting measurements across the device. Fig.~\ref{fig:fig1}(e) shows calculations of the cross capacitance---the relative gate-to-dot capacitance between a second gate and the dot's plunger gate---for the outer barrier gate (e.g., gates B1 and B3 for dots under P2 and P3, respectively, in the P2-P3 double dot). The capacitance model, which treats the electron as a rectangular distribution of charge and does not account for any disorder-induced variations of the potential or heterostructure, allows extraction of the dot center position in the channel via Fig.~\ref{fig:fig1}(f) with \SI{2}{\nano\meter} resolution.

\subsection{Valley spectroscopy\label{subsec:spectroscopy}}
\begin{figure}
    \centering
    \includegraphics{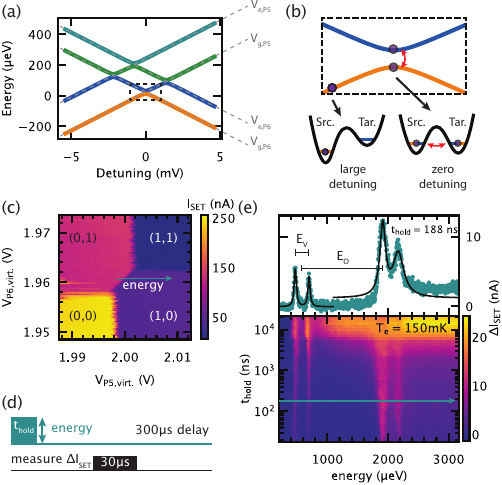}
    \caption{\bf Detuning axis pulsed spectroscopy (DAPS) valley splitting measurement. \rm (a) Single-electron energy level diagram for the example P5-P6 double dot, with anti-crossings arising between ground and excited valley states. (b) When the electron (purple circle) is far-detuned from an anti-crossing it sits in the ground valley of the DAPS source (Src.) dot. At an anti-crossing the valley states hybridize and the electron can tunnel into DAPS target (Tar.) dot. (c) Double dot charge stability diagram focused on the single electron polarization line. Teal arrow shows the direction of the voltage pulse in (d) and (e). (d) DAPS measurement diagram with a variable-time ($t_{hold}$), variable-energy voltage pulse followed by a \SI{30}{\micro\second} SET current measurement and a \SI{300}{\micro\second} delay to allow the electron to relax back to the source dot. (e) DAPS spectrum of the P5 dot at an electron temperature of $T_e=\SI{150}{\milli\kelvin}$, revealing valley-orbital states. Valley splitting $E_V$ is extracted from fitting the two lowest peaks in line cuts at one $t_{hold}$ value, and orbital energy $E_O$ is extracted from fitting the first excited orbital (top).}
    \label{fig:fig2}
\end{figure}

All measurements in this work are performed with the device tuned into a double quantum dot (DQD) regime, with all other qubit-side finger gates set at or near \SI{1}{\volt} and the unused SET gates set at \SI{0.2}{\volt}. The measurement setup is described in SI Sec.~S2. In Fig.~\ref{fig:fig2} we describe how DAPS reveals the valley-orbital spectrum, enabling measurement of valley splitting energy $E_V$ and dot radius, using the P5-P6 DQD as an example. Fig.~\ref{fig:fig2}(a) shows the calculated energy level diagram for the four lowest-lying states in the DQD, found by diagonalizing the Hamiltonian with an \textit{a priori} valley splitting and tunnel couplings between each state. The ground valleys, $V_{g,i}$, and the excited valleys, $V_{e,i}$, in each dot $i$ are shown for measured P5 and P6 valley splittings, with an artificially large tunnel coupling to accentuate the anti-crossings between each state.

Fig.~\ref{fig:fig2}(c) shows a charge stability diagram for the P5-P6 DQD focused on the single electron polarization line, measured via the SET current $I_{SET}$. Plunger gates are virtualized to each other and to the relevant barrier gates (B4, B5, and B6, here). The beginning of latching is visible at the P5 charging lines, indicating low coupling to the left reservoir. As the DAPS measurement here does not rely on coupling to a reservoir, pinching off this coupling ensures the resulting signal is only from the electron tunneling between the two dots. The teal arrow shows the direction of the pulse applied in the measurement diagram in Fig.~\ref{fig:fig2}(d), where a variable energy, variable time ($t_{hold}$) voltage pulse detunes the double dot potential, followed by a measurement of $I_{SET}$. After the current measurement, a \SI{300}{\micro\second} delay time allows the electron to relax from the target dot level (an excited state at the initial/final detuning) back to the source dot ground state.

The full DAPS spectrum versus pulse energy and $t_{hold}$ for DQD P5-P6 is shown in Fig.~\ref{fig:fig2}(e), with a linecut at $t_{hold}=\SI{188}{\nano\second}$ shown on top. The electron temperature $T_e=\SI{150}{\milli\kelvin}$ is measured from the single electron charging linewidth, and the fridge mixing chamber plate sits at $T_{mxc}=\SI{35}{\milli\kelvin}$. All measurements in this work are performed at this temperature. The spectrum shows two well-resolved orbital states, each with a ground and excited valley state. Fitting the ground orbital peaks with two lorentzian curves we extract the magnitude of the dot $E_V$, taking the linewidth as the error, following Ref.~\cite{chen_detuning_2021}. The higher-lying peaks near \SI{2000}{\micro\eV} are the ground and excited valleys in the first excited $x$-orbital, at energies $E_O\pm E_{V,e}/2$ relative to the middle of the ground orbital valley energies. The dot radius along the dot channel is then calculated as $r_x=\sqrt{2\hbar^2/(m_e^* E_O)}$, where $m_e^*=0.19m_e$ is the transverse effective electron mass in Si. Across all dots in this device, it is not always possible to resolve the two excited-orbital valley states, likely due to variation in dephasing and tunneling rates. For consistency across measurements, we calculate $E_O$ by approximating the excited orbital energy as the first peak in the excited orbital spectrum (around \SI{1950}{\micro\eV} in Fig.~\ref{fig:fig2}(e)) and take the ground orbital energy as the average of the two valley peaks in the ground state spectrum (as marked in Fig.~\ref{fig:fig2}(e)). As this excited orbital energy corresponds to the lower valley state, this method biases $E_O$ down by $\approx\SIrange{50}{100}{\micro\eV}$, based on resolvable excited-orbital valley splitting energies. This does not affect our conclusions related to the dot size. Energies are calculated based on the lever arm measured under each plunger/barrier finger gate, described in SI Sec.~S3.

\subsection{Continuous valley splitting correlations\label{subsec:valleyprobe}}
\begin{figure*}
    \centering
    \includegraphics[width=\textwidth]{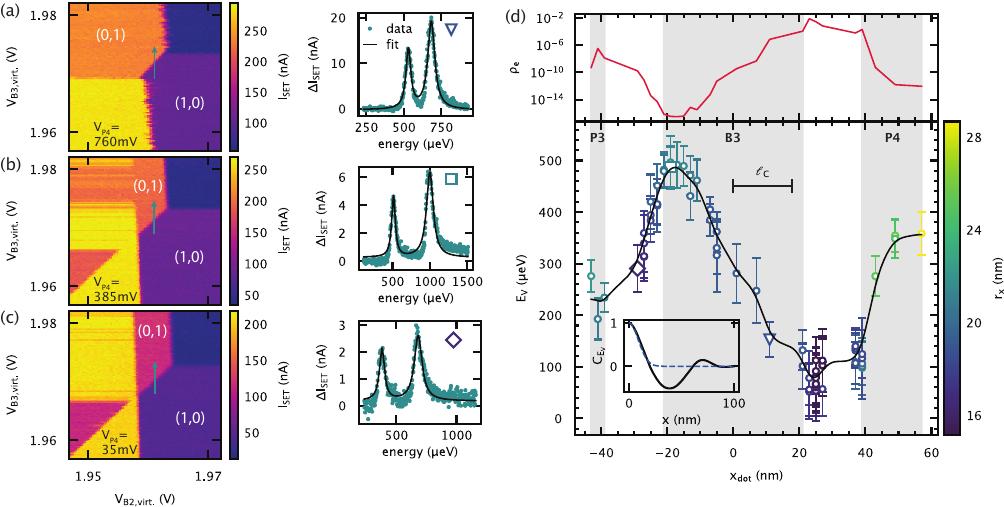}
    \caption{\bf Continuous electron valley probe. \rm (a-c) Charge stability diagrams (left) of B2-B3 double dot single electron polarization line and corresponding DAPS measurements (right) as the B3 dot's outer barrier (P4) is depleted. DAPS measurements are fit with double-Lorentzians to extract valley splitting energy ($E_V$). Teal arrows show the direction of the DAPS voltage pulse. Triangle, square, and diamond markers in the DAPS plot correspond to the three points in (d). (d) $E_V$ versus center position of the electron $x_{dot}$ accumulated under P3, B3, and P4 dots, relative to the center of gate B3, where error is calculated from the DAPS peak linewidth~\cite{chen_detuning_2021}. An interpolated curve with \SI{2}{\nano\meter} gaussian smoothing is overlaid. Inset is the autocorrelation $C_{E_V}$ (solid black) calculated from the interpolated curve and the fit (dashed blue) from theory. The scale bar is the correlation length from the inset figure. On top is a calculation of the thermal population $\rho_e$ at \SI{150}{\milli\kelvin} in the excited valley state for each measured $E_V$. Electron radii are calculated from the orbital energy extracted from DAPS.}
    \label{fig:fig3}
\end{figure*}

We now employ DAPS as a local probe of correlations in the valley splitting energy, focusing on the P3-B3-P4 region. Rather than treat the valley splitting as a property of the quantum dot, we translate the target dot potential in space such that the valley splitting value continuously probes the underlying material. In Fig.~\ref{fig:fig3}(a-c) we show charge stability diagrams and DAPS valley spectra for the B2-B3 DQD as we deplete the P4 outer barrier voltage $V_{P4}$ (the roles of the ``B'' and ``P'' gates are swapped here) from an initial regime at $V_{P4}=\SI{760}{\milli\volt}$ where the B3 dot is coupled to the reservoir down to $V_{P4}=\SI{35}{\milli\volt}$ where there is visible latching in the (0,0) charge state. However, we are still able to pulse across the polarization line (teal arrows) and can still perform DAPS. As P4 is depleted we measure the valley and orbital energies as described in Sec.~\ref{subsec:spectroscopy}, and we calculate the $x$-axis dot location $x_{dot}$ from a cross-capacitance measurement to gate P4 and the simulations in Fig.~\ref{fig:fig1}(e-f).

We perform further measurements on the P3-P4 DQD, with cross-capacitance measured to B2 (B4) for dot P3 (P4), and combine with the B2-B3 data to generate the continuous valley probe in Fig.~\ref{fig:fig3}(d) over \SI{100}{\nano\meter} of material, where the colorscale encodes the dot radius at each point. The position $x_{dot}$ is relative to the center of gate B3. We observe fast fluctuations in the valley splitting under single finger gates from a maximum value of \SI{500\pm50}{\micro\eV} at \SI{-19}{\nano\meter}, near the left edge of the B3 gate, to a minimum value of \SI{50\pm50}{\micro\eV} at \SI{23}{\nano\meter}, near the right edge of the B3 gate. In SI Sec.~S4 we instead plot $E_V$ as a function of $r_x$, finding local trends but no overall behavior, suggesting that the observed variation is predominantly from the underlying material disorder.

At the top of Fig.~\ref{fig:fig3}(d) we calculate the thermal population of the excited valley state $\rho_e$ at an electron temperature of \SI{150}{\milli\kelvin}, demonstrating the sensitivity of this population on $E_V$. Notably, in the regions where $E_V<\SI{60}{\micro\eV}$, $\rho_e\approx\SI{1}{\%}$, which may impact Pauli spin blockade fidelity~\cite{tagliaferri_impact_2018}. This highlights the importance of engineering uniformly large $E_V$ to avoid local hot-spots that will impact qubit operation, including during electron shuttling and Pauli spin blockade~\cite{volmer_mapping_2024,losert_strategies_2024,nemeth_omnidirectional_2025}.

We overlay an interpolated curve on Fig.~\ref{fig:fig3}(d) with a \SI{2}{\nano\meter} gaussian filter from which we calculate the autocorrelation of the valley splitting $C_{E_V}$, shown in the inset. This curve crosses the $x$-axis as a result of sampling a finite region of space. We fit this to the theoretical model~\cite{losert_practical_2023,volmer_mapping_2024}, shown in dashed blue and derived in Methods,
\begin{equation}\label{eq:CEV}
    C_{E_V}(x)=\exp\left(-\frac{2}{4-\pi}\frac{x^2}{\ell_C^2}\right)
\end{equation}
where $\ell_C$ is the correlation length. This reveals a correlation length of \SI{19.2}{\nano\meter}, shown in the scale bar, matching the average dot radius of \SI{19}{\nano\meter} in this dataset. Thus the observed $E_V$ fluctuations arise from the electron wavefunction smoothing over faster variations in the underlying material, and the resolution of this probe is limited by the dot radius, consistent with the same physical mechanism in Ref.~\cite{volmer_mapping_2024}. In SI Sec.~S5 we show how this analysis depends on the gaussian filter size.

\begin{figure}
    \centering
    \includegraphics{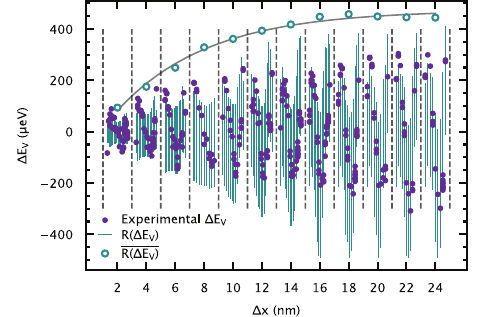}
    \caption{\bf Expected $E_V$ variations. \rm The valley splitting difference $\Delta E_V$ (purple filled in circles) is plotted alongside the 5\%--95\% range $R(\Delta E_V)$ of valley splitting at some distance $\Delta x$ (teal lines). The mean expected range for each distance $\overline{R(\Delta E_V)}$ is plotted atop the data (open teal circles), fit to an exponential decay (gray line).}
    \label{fig:fig4}
\end{figure}

We now corroborate the observed variations in $E_V$ by modeling the expected range in the presence of alloy disorder. Given a known $E_V$ at some dot position, we use a joint probability density function to predict the range of $E_V$ when forming a dot some distance away. The variance $\sigma_\Delta^2$ of the complex inter-valley coupling $\Delta$, where $E_V=2\abs{\Delta}$, at some point in space is calculated assuming $\overline{E_V} = \sqrt{\pi} \sigma_\Delta$, as predicted in the alloy disorder-dominated regime~\cite{losert_practical_2023}, and a dot size given by the average measured orbital energy. More details on the calculation can be found in Methods.

In Fig.~\ref{fig:fig4} we plot the valley splitting difference $\Delta E_V$ between any two points from Fig.~\ref{fig:fig3}(d) at a distance $\Delta x$ away. As the dot position simulations in Fig.~\ref{fig:fig1}(e,f) have a \SI{2}{\nano\meter} resolution, all $\Delta x$ values are in \SI{2}{\nano\meter} intervals and datapoints are only offset in $x$ for clarity. Then we calculate with Eq.~\eqref{eq:P} the range of valley splitting $R(\Delta E_V)$ we would expect for the right (more positive) dot position given the measured value at the left (more negative) dot position, shown in the vertical teal lines. The increase in $R(\Delta E_V)$ tracks with the increase in experimentally measured differences. The open teal circles are the average expected range $\overline{R(\Delta E_V)}$. Fitting to a simple exponential decay, we find that the magnitude of the range of these predictions decays over \SI{6.4\pm.6}{\nano\meter}. $\overline{R(\Delta E_V)}$ saturates as $\Delta x$ approaches the correlation length $\ell_C$ calculated above, clarifying how local correlations arising from alloy disorder die out on or below the lengthscale of the electron wavefunction size.

\subsection{Device-scale valley splitting and correlations}

\begin{figure}
    \centering
    \includegraphics{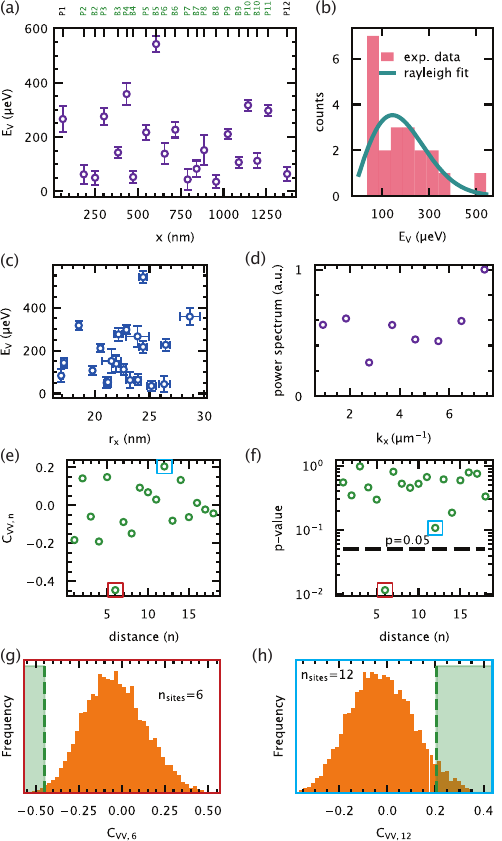}
    \caption{\bf Valley splitting in full array. \rm (a) Valley splitting $E_V$ measured in 21 separate dots in the Tunnel Falls device, with positions extracted from the capacitance model in Fig.~\ref{fig:fig1}(e). (b) $E_V$ histogram revealing deviations from theoretical Rayleigh distribution, consistent with finite random sampling. (c) $E_V$ versus electron radius along the channel axis, showing no overall dependence. Error bars in (a) and (c) are calculated from the DAPS peak linewidth~\cite{chen_detuning_2021}. (d) Fourier transform of $E_V$ along the array, with no components consistent with long-range correlations. (e) Autocorrelation $C_{VV,n}$ versus distance $n$ with large negative $C_{VV}$ at $n=6$ and moderate positive $C_{VV}$ at $n=12$. (f) Statistical significance (p-value) of observed $C_{VV,n}$ values calculated with a permutation test. Null distributions for permutation tests on $n=6,12$ are shown in (g,h), respectively, with shaded area showing the calculated p-values from (f).}
    \label{fig:fig5}
\end{figure}

We now probe variation in $E_V$ across the \SI{1.3}{\micro\meter} device channel in Fig.~\ref{fig:fig5}(a), sampling under each accessible finger gate (as shown in Fig.~\ref{fig:fig1}(a)). The finger gate pitch is greater than the extracted correlation length, thus we initially treat the values as uncorrelated and fit the histogram of these data, in Fig.~\ref{fig:fig5}(b), to the theoretically expected Rayleigh distribution. We find a mean valley splitting $\overline{E_V}=\SI{179}{\micro\eV}$ and a maximum value of \SI{540\pm30}{\micro\eV}. While this dataset is weighted towards $E_V<\SI{100}{\micro\eV}$, skewing the Rayleigh fit, we show in SI Sec.~S6 that this is within the expected behavior for 21 randomly drawn samples from the simulated ADD data from Fig.~\ref{fig:fig1}(d). In Fig.~\ref{fig:fig5}(c) we find no significant dependence of $E_V$ on the dot radius, concluding as in Sec.~\ref{subsec:valleyprobe} that the underlying material disorder dominates the observed variation. In Fig.~\ref{fig:fig5}(d) we plot the Fourier transform of the spatially varying $E_V$ data, finding no notable oscillatory behavior aside from large nearest-neighbor variation indicated by the value at the highest $k$-vector.

\begin{figure}
\centering
\includegraphics{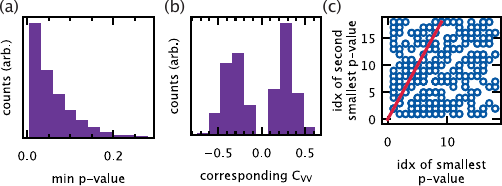}
\caption{\bf Simulated permutation test. \rm (a) Distribution of lowest p-values found in autocorrelation permutation tests run on randomly sampled 19-point datasets from simulated data in Fig.~\ref{fig:fig1}(c). (b) Distribution of $C_{VV}$ values corresponding to p-values in (a). (c) Scatter plot of simulated datasets matching the observed correlations in Fig.~\ref{fig:fig5}(e,f) where the most-significant $C_{VV}<0$ and the second-most-significant $C_{VV}>0$. The red line shows datasets where $n(C_{VV}>0)$ is twice $n(C_{VV})<0$, as in the measured dataset.}
\label{fig:fig6}
\end{figure}

\section{Discussion}
Device-scale correlations in $E_V$ values would have implications for studies of the underlying material disorder as well as correlations in errors between qubits in a single device. To investigate correlations we approximate the middle 19 dots [shown in green in Fig.~\ref{fig:fig5}(a)] as evenly spaced and calculate the normalized autocorrelation $C_{VV,n}$ between the valley splitting $n$ number of gates apart, shown in Fig.~\ref{fig:fig5}(e). The largest magnitude correlations are found at distances $n=6$ (negative, anti-correlation) and $n=12$ (positive correlation), suggesting an overall correlation length of \SI{720}{\nano\meter} not visible in Fig.~\ref{fig:fig5}(d).

We evaluate the significance of these correlations with a permutation test, with extracted p-values shown in Fig.~\ref{fig:fig5}(f). In a permutation test, the given dataset is rearranged into many permutations and the statistic---here the autocorrelation $C_{VV,n}$---is calculated, generating a null distribution, shown for $n=6,12$ in Fig.~\ref{fig:fig5}(g,h), respectively. The permutation test then calculates a p-value based on where in the null distribution the statistic for the real data lies, shown with the green dotted line and shaded regions in Fig.~\ref{fig:fig5}(g,h). Comparing to a p-value of 0.05, a standard for statistical significance, we find that the anti-correlations at $n=6$ are more significant than the correlations at $n=12$, which can arise due to the finite sampled length of the data. The permutation test demonstrates that on the intermediate device scale the fluctuations in $E_V$ between spatially separated dots display correlations.

In Fig.~\ref{fig:fig6} we investigate whether the observed correlations are consistent with randomly sampling an underlying Rayleigh distribution of $E_V$. Starting from the calculated values in Fig.~\ref{fig:fig1}(d), we generate many ordered sets of 19 values, consistent with the analysis in Fig.~\ref{fig:fig5}, and perform the same autocorrelation permutation test. In Fig.~\ref{fig:fig6}(a,b) we plot the histograms of the lowest observed p-value and the corresponding $C_{VV}$, respectively. We find that it is common to observe large, significant correlations or anti-correlations in a given dataset, similar to the measured dataset. In Fig.~\ref{fig:fig6}(c) we specifically find that it is possible to observe the same situation in Fig.~\ref{fig:fig5}(e-f), with the lowest p-value occurring for $C_{VV}<0$ and the second-lowest p-value occurring for $C_{VV}>0$. The blue circles plot the correlation distances for these situations. The red line specifically shows the situations where $n(C_{VV}>0)=2\cdot n(C_{VV}<0)$, mimicking our observations. As this can arise from randomly sampling the simulated $E_V$ distribution, we conclude that the observed correlations are consistent with a single ADD distribution describing the behavior in the entire device. Our results and analysis demonstrate that, in finite-sampled regions of material, fluctuations that appear as correlations between nearby sites are expected. Uncovering any large-scale variation in the underlying material that would cause correlated behavior in scaled-up devices will require larger datasets on larger devices, perhaps coupled with nanoscale probes of material properties, as achieved through x-ray microscopy~\cite{evans_nanoscale_2012,holt_strain_2014,zoellner_imaging_2015,park_electrode-stress-induced_2016,hruszkewycz_high-resolution_2017,pateras_mesoscopic_2018,corley-wiciak_nanoscale_2023}.

By probing valley splitting both continuously under neighboring gates and discretely across 21 gates of a single 2.8\%-Ge Si quantum well device we have observed the impact of inherent material inhomogeneity on the emergent properties of quantum electronic devices. We verified that in state-of-the-art, industrially fabricated devices the valley splitting behaves according to the ADD theory. By analyzing the nanometer-scale correlations between valley splitting measurements we found local decorrelation on a \SI{19}{\nano\meter} lengthscale, smaller than the size of the electron wavefunction, demonstrating that as the dot potential shifts the electron it samples a new region of material uncorrelated with the prior region. On the scale of the entire \SI{1.3}{\micro\meter} device we revealed device-scale correlations in valley splitting arising from fluctuations consistent with random sampling of the ADD theoretical distributions.

Low-lying valley excited states, arising in Si quantum dots from the underlying band structure and material variation, stand to limit the scalability of Si-based quantum computers. In addition to predicting and engineering the average valley splitting properties in quantum wells at wafer-scale, we find that it is also critical to understand how local material variation manifests at single-gate and device scales. Ultimately, developing robust modeling and local probes, as well as generating larger experimental datasets, will inform future developments of scalable quantum devices by providing a window into the effects of the underlying material host.

\section{Methods}

\subsection{Valley splitting simulations}

To generate the valley splitting histogram of Fig.~\ref{fig:fig1}(d), we use the two-band tight-binding model of Boykin et al.~\cite{boykin_valley_2004}, where nearest-neighbor and next-nearest-neighbor hopping parameters in the $z$ direction are chosen to match the position and curvature of the valley minimum in the Si bandstructure.
Following Refs.~\cite{paquelet_wuetz_atomic_2022, losert_practical_2023}, we employ a 1D model with onsite potential $U = U_\text{qw} + e E_z z$, where $U_\text{qw}$ is the quantum well potential and $E_z = 1$~mV/nm is the vertical electric field, consistent with recent experiments \cite{hollmann_large_2020}.
The quantum well potential is given by
\begin{equation}
U_\text{qw}(z_l) = \Delta E_c \frac{ Y_l }{ Y_s - Y_w }   
\end{equation}
where $Y_l$ is the Ge concentration in layer $l$, $Y_w = 2.8$\% is the Ge concentration in the quantum well, $Y_s = 30$\% is the Ge concentration in the SiGe barrier (substrate) region, and $\Delta E_c = 139$~meV is the conduction band offset, computed for this quantum well following Refs.~\cite{paquelet_wuetz_atomic_2022, losert_practical_2023}. 
We assume the quantum well interfaces are sigmoidal, with an average profile given by
\begin{equation}
    \bar Y_l = Y_w + \frac{Y_s - Y_w}{1 + \exp[(z - z_t)/\tau]} + \frac{Y_s - Y_w}{1 + \exp[(z_b - z)/\tau]}
\end{equation}
where we assume the interface width $\lambda = 4\tau =1$~nm and the quantum well width $z_b - z_t = 4.6$~nm.
To account for the disorder due to random fluctuations in the composition of the SiGe alloy, we randomize $Y_l$ slightly at each layer by adding a small fluctuation $\delta Y_l$, where $\delta Y_l$ is sampled according to an effective binomial distribution \cite{paquelet_wuetz_atomic_2022, losert_practical_2023}
\begin{equation}
    Y_l \sim \frac{1}{N_\text{eff}}\text{Binom}(N_\text{eff}, \bar Y_l)
\end{equation}
where $N_\text{eff} = 2 \pi r^2 / a_0^2$, $r = \sqrt{2 \pi / \hbar E_O}$, $E_O = 2$~meV, and $a_0=\SI{0.543}{\nano\meter}$ is the lattice constant of Si.
We repeat this randomization procedure 1000 times, to build up a distribution of valley splittings. 
Finally, to account for multi-band effects, we scale these valley splittings by a correction factor of 75\%, known from comparisons to more sophisticated atomistic tight-binding models \cite{paquelet_wuetz_atomic_2022, losert_practical_2023}.

\subsection{Dot-scale $E_V$ correlations}

Now, we derive the formula for the autocorrelation curve of the valley splitting, $C_{E_V}$. We define $E_V = 2|\Delta|$, where $\Delta$ is the inter-valley coupling, and $\Delta = \delta e^{i \phi}$, where $\delta = |\Delta| = E_V / 2$. We start with the joint probability density function for two measurements, $\delta_1$ and $\delta_2$, evaluated at a distance $d$ apart on the heterostructure~\cite{losert_practical_2023,klos_atomistic_2024}:
\begin{multline} \label{eq:joint_pdf_delta_mag}
    p(\delta_1, \delta_2; d) = \frac{4 \delta_1 \delta_2}{\sigma_\Delta^2} \left(1 + e^{-4 d^2 / r^2} - 2 e^{-2 d^2 / r^2} \right)^{1/2} \\
    \exp \left[ - \frac{(\delta_1^2 + \delta_2^2)e^{2 d^2 / r^2}}{\left(e^{2 d^2 / r^2} - 1 \right) \sigma_\Delta^2} \right] I_0 \left[ \frac{\delta_1 \delta_2}{\sigma_\Delta^2} \text{csch}\left( \frac{d^2}{r^2} \right) \right]
\end{multline}
where $r = \sqrt{2 \hbar^2 / m_e^* E_O}$.
For simplicity, we write Eq.~\eqref{eq:joint_pdf_delta_mag} for isotropic dots ($r = r_x = r_y$), but the anisotropic case can be found by replacing everywhere $d^2 / r^2 \rightarrow d_x^2 / r_x^2 + d_y^2 / r_y^2$. We use the definition of covariance, $\text{Cov}[\delta_1, \delta_2] = \langle \delta_1 \delta_2 \rangle - \langle \delta_1 \rangle \langle \delta_2 \rangle$, where $\langle \cdot \rangle$ denotes an expectation value. First, 
\begin{multline}
    \langle \delta_1 \delta_2 \rangle = \int_0^\infty d\delta_1 \; d\delta_2 \; \delta_1 \delta_2 \; p(\delta_1, \delta_2; d) \\
    = \frac{\sigma_\Delta^2}{2} \left\{2 E\left[e^{-\frac{2d^2}{r^2}} \right] - \left[1 - e^{-\frac{2 d^2}{r^2}} \right] K\left[e^{-\frac{2 d^2}{r^2}} \right] \right\}
\end{multline}
where $K(m)$ and $E(m)$ denote complete elliptic integrals of the first and second kind. Since $\Delta$ is complex Gaussian with zero mean, $\delta$ has a Rayleigh distribution, with mean $\langle \delta \rangle = \sigma_\Delta \sqrt{\pi} / 2$ and variance $\langle \delta^2 \rangle = (1 - \pi/4) \sigma_\Delta^2$. Substituting $E_V = 2|\delta|$, and using the definition of the correlation function $C_{E_V}(d) = \text{Cov}[E_{V1}, E_{V2}; d] / \sigma_{E_V}^2$, where $\sigma_{E_V}^2 = \langle (E_V - \langle E_V\rangle)^2 \rangle = (4 - \pi)\sigma_\Delta^2$, we have the correlation function
\begin{multline} \label{eq:exact_Ev_corr}
    C_{E_V}(d) = \\
    \frac{2}{4 - \pi} \left\{ 2 E\left[e^{-\frac{2d^2}{r^2}} \right] - \left[ 1 - e^{-\frac{2 d^2}{r^2}} \right]K\left[e^{-\frac{2 d^2}{r^2}} \right] \right\} \\
    - \frac{\pi}{4 - \pi}.
\end{multline}
Keeping terms less than $O(x^2)$, the series expansions of $E[\exp[-x]]$ and $K[\exp[-x]]$ are 
\begin{equation} \label{eq:series_exp}
\begin{split}
    E[\exp[-x]] &\approx 1 + \frac{1}{4} \left(4 \ln 2 - 1 - \ln(x) \right)x \\
    K[\exp[-x]] &\approx 2 \ln 2 - \frac{1}{2} \ln x + \frac{1}{8} \left(4 \ln 2 - \ln x \right)x. \\
\end{split}
\end{equation}
Using Eq.~\eqref{eq:series_exp} in Eq.~\eqref{eq:exact_Ev_corr}, and keeping terms to $O(d^2)$, we can approximate Eq.~\eqref{eq:exact_Ev_corr} as a Gaussian, resulting in the much simpler expression 
\begin{equation} \label{eq:approx_Ev_corr}
    C_{E_V}(d) \approx 1 - \frac{2}{4 - \pi}\frac{d^2}{r^2} \approx \exp \left[ -\frac{2}{4 - \pi} \frac{d^2}{r^2} \right].
\end{equation}
Despite applying a series expansion in small $d/r$, we find Eq.~\eqref{eq:approx_Ev_corr} is a good approximation over the full range of $d$.

\subsection{Valley splitting predictions using correlations~\label{app:predcorr}}

Given a known valley splitting, we can use Eq. \eqref{eq:joint_pdf_delta_mag} to predict the valley splitting some distance $d$ away. We will take two sites with known orbital splitting, then use a measured valley splitting at one site to predict a range of expected valley splitting values at the other site. First, we use orbital splitting values $E_{O_1}, E_{O_2}$ to approximate $\sigma_{\Delta_1}, \sigma_{\Delta_2}$ at each site using
\begin{equation}\label{eq:sD}
    \sigma_\Delta^2 = \eta_z \int dx \,dy\, |\psi(x,y)|^4
\end{equation}
where $\psi(x,y)$ is the electron wavefunction in the plane of the quantum well and $\eta_z$ is a constant describing alloy disorder in the $z$ direction transverse to the well plane~\cite{losert_2025}. As the $y$ orbital confinement is not known in most cases, but assumed to be separable from the $x$ confinement, we can approximately replace  Eq.~(\ref{eq:sD}) by
\begin{equation}\label{eq:sD2}
    \sigma_\Delta^2 = \tilde\eta_z \int dx\, |\psi(x)|^4 .
\end{equation}
To approximate $\tilde\eta_z$, we  solve Eq. \eqref{eq:sD2} using the mean $x$-orbital splitting to estimate $\psi(x)$, assuming a harmonic confinement potential, and we use the mean valley splitting value to estimate $\sigma_\Delta$, assuming $\overline{E_V} = \sqrt{\pi} \sigma_\Delta$, as predicted in the alloy disorder-dominated regime~\cite{losert_practical_2023}. Using this value of $\tilde\eta_z=\SI{5.38e-4}{\micro\eV^2m^2}$, we can predict $\sigma_\Delta$ at each site to higher accuracy compared to using a sample-averaged value.

. In the joint probability density function, we will take $E_O = \frac{1}{2} (E_{O_1} + E_{O_2})$ and $\sigma_\Delta = \frac{1}{2} (\sigma_{\Delta_1} + \sigma_{\Delta_2})$. With our known value of $\delta_1$, we can normalize the joint probability density function by

\begin{equation}
    N = \int_0^\infty d\delta_2 p(\delta_1  =\delta_1, \delta_2; d)
\end{equation}

so the probability of observing some $\delta_2$ in the range $[a,b]$ becomes

\begin{equation}
\label{eq:P}
    P(\delta_2;a,b) = \frac{1}{N} \int_a^b d \delta_2 p(\delta_1  =\delta_1, \delta_2; d)
\end{equation}

To compute the $5^{\text{th}}$ and $95^{\text{th}}$ percentiles, we optimize lower and upper thresholds $T_a, T_b$ such that $P(\delta_2; 0, T_a) = 0.05$ and $P(\delta_2; T_b, \infty) = 0.05$.

In the case of many measurements at a single site $i$, we take $\delta_i, E_{O_i}$ to be the mean of all measurements performed at that site. $\sigma_{\Delta_i}$ is then computed using the mean value of $E_{O_i}$.

\section{Data availability}
Data and analysis for this work will be made available upon request.

\section{Acknowledgments}
 This material is based upon work supported by the U.S. Department of Energy Office of Science National Quantum Information Science Research Centers as part of the Q-NEXT center (J.C.M., J.R., J.Z., F.J.H., M.A.E.). Modeling and simulation work was sponsored by the Army Research Office under Awards No. W911NF-23-1-0110 and W911NF-22-1-0090 (E.E., M.P.L., T.O., M.F.) and the Oak Ridge Institute for Science and Education (ORISE) (E.B.). Measurement and analysis assistance were supported by the Grainger Fellowship from the University of Chicago (C.S.W.). The authors acknowledge Intel Corp. for providing the quantum computer system and engineering support. This work benefited from discussion with the University of Chicago Department of Statistics Consulting Program. The views, conclusions, and recommendations contained in this document are those of the authors and are not necessarily endorsed by nor should they be interpreted as representing the official policies, either expressed or implied, of the Army Research Office or the U.S. Government. The U.S. Government is authorized to reproduce and distribute reprints for U.S. Government purposes notwithstanding any copyright notation herein.

\section{Author Contributions Statement}
J.C.M., M.A.E., J.R., and M.F. conceived of the study; J.C.M. performed measurements and coordinated the project; E.E., E.C.B., M.P.L., T.O., and F.A.M. performed theoretical calculations; J.C.M. analyzed the data with assistance from J.R., C.S.W., M.A.E., and M.F.;  J.R., C.S.W., D.K., F.L., M.J.C., J.Z., and F.J.H. assisted in measurements; F.L. built the measurement software framework; D.K., F.L., and M.J.C. developed measurement protocols and provided Intel hardware and software; Intel Corp. provided the measured Tunnel Falls device. J.C.M., E.E., E.C.B., and M.P.L. prepared the manuscript with input from all authors.

\section{Competing Interests Statement}
M.P.L and M.F. have applied for a patent on the Ge-containing quantum well structure described here. The remaining authors declare no competing interests.

\end{document}